\documentclass[a4paper]{article}
  \usepackage{ graphicx }

  \newlength\semitextwidth
  \newlength\xx
  \advance\textwidth by -2in

\makeatletter
  \def\@biblabel#1{}
  \def\@cite#1{#1}
\makeatother

\def\affil#1{
\begin{center} 
  #1
\end{center}
}
\date{(\today)}
  \def\nroot{n_{\rm FOP}}
  
  \font\bfit=cmbxti10
  \def\VEC#1{\hbox{\bfit #1}} 
  \def\IMG{.}  
  \def\EXT{eps}
  \def\Null#1{}
  \def\del#1#2{\frac{\partial #1}{\partial #2}}
  \def\d{\partial}
  \def\={\fallingdotseq}
  \def\reg#1{{\rm reg}(#1)}
  \def\pitem#1#2#3#4{%
    \leavevmode\hbox to 114mm{%
    (#1) $(a,b)=(#2)$\hfill$(\nroot,\nroot^*)=(#3)$}\par%
    \includegraphics{\IMG/#4}\par%
  }
  \def\pit#1#2#3{\leavevmode\hbox to 114mm{(#1) $(a,b)=(#2)$\hfill}\par\includegraphics[width=4.5in, height=2.5in]{\IMG/#3}\par}
  \makeatletter
  \def\backfigure{\advance\c@figure by -1}
  \makeatother

  \def\fallingdotseq{
     \hbox{%
     \kern0.23em\raisebox{0.48em}{.}\kern-0.23em%
     =\kern-0.28em\raisebox{-0.1em}{.}\kern0.28em%
     }%
   }
\begin{document}
\def\emaila{\footnote{sasame2005@mail.goo.ne.jp}}
\def\emailb{\footnote{tanikawa.ky@nao.ac.jp}}
\title{\bf \Large The Rectilinar Three-body Problem using Symbol Sequence \\
II. Role of the periodic orbits}
\author{$^{1,2}$\emaila\ Masaya Masayoshi Saito 
 and $^{1,2}$\emailb\ Kiyotaka Tanikawa}

\maketitle

\affil{$^1$Department of Astronomical Science, SOKENDAI the graduate university,\\
Shonan International Village, Hayama, Kanagawa 240-0193, JAPAN}

\affil{$^2$Division of Theoretical Astronomy, National Astronical Observatory of Japan, \\
Osawa 2-21-1, Mikata, Tokyo 181-8588, JAPAN}

\begin{abstract}
  We study the change of phase space structure of the rectilinear three-body
problem when the mass combination is changed. Generally, periodic orbits 
bifurcate from the stable Schubart periodic orbit and move radially outward. 
Among these periodic orbits there are {\it dominant} periodic orbits having 
rotation number $(n-2)/n$ with $n\geq3$. We find that the number of 
dominant periodic orbits is two when $n$ is odd and four when $n$ is even.
Dominant periodic orbits have large stable regions in and 
out of the stability region of the Schubart orbit (Schubart region), 
and so they determine the size of the Schubart region and influence 
the structure of the Poincar\'e section out of the Schubart
region. Indeed, with the movement of the dominant periodic orbits, 
part of complicated structure of the Poincar\'e section follow these orbits. 
We find stable periodic orbits which do not bifurcate from the 
Schubart orbit.
\end{abstract}

\section{Introduction}
  In the present study, we continue our work on the rectilinear 
three-body problem (Saito \& Tanikawa, 2007; hereafter 
referred to as Paper I). The rectilinear three-body system is such that 
three particles are on a line. This problem has two degrees of freedom.
\par
The structure on the surface of section of the rectilinear three-body 
system has been studied using
the Poincar\'e map (\cite{HM1993}; hereafter HM1993).
They found that the Poincar\'e section is divided into
three basic regions: the Schubart region, the chaotic scattering 
region, and the immediate escape region. Inside the scattering region, 
the interplay time is so sensitive to the initial conditions 
that there seemed to be no structure.
HM1993 also found that the number of scallops, which constitute the
immediate escape region, increases as the mass of the central particle 
becomes smaller.
\par
Tanikawa and Mikkola (2000a; hereafter referred to as TM2000a) studied the 
structure of the Poincar\'e section for the equal-mass case using 
symbol sequences which record the collisional history of orbits. 
They used the Poincar\'e section as the initial condition 
surface and associated with each point of the surface its future history 
including the final motion. They then succeeded in showing that 
the chaotic scattering region is filled with points whose orbits end  
with triple collision and that these points form well stratified curves.
TM2000a's fine structure of the Poincar\'e section for the equal-mass 
case shows that (1) stratified triple collision curves make a 
{\it sector} together with the sub-region of the immediate escape region 
(i.e., scallop); (2) four sectors surround the Schubart region. 
\par
The study of a similar dynamical system,
the collinear Coulomb three-body problem with electron-ion-electron
configuration, carried out by Sano (2003) is helpful to understand
our system. In his system, there exist the Schubart region and 
unstable periodic points on its vertices. He confirmed the existence
of the separatrices of these unstable points in his system numerically 
following the mapped points starting near the vertices.
\par
In Paper I, we studied how the structure on the Poincar\'e section 
changes as the central mass varies. The chaotic scattering region 
is foliated with triple collision curves in general mass combination. 
This foliation makes two types of block: arch- and germ-shaped blocks 
(hereafter simply arches and germs; see Fig. \ref{struct00}).  
The chaotic scattering region consists only of arches near the mass 
combinations of total degeneracy, whereas it consists of arches and 
germs for the mass combinations away from total degeneracy.  
Here by total degeneracy we mean that stable and unstable manifolds of 
the two fixed points on the Triple Collision Manifold connect smoothly 
(McGehee, 1974). 
In the totally degenerate cases, the solutions can be analytically continued 
beyond triple collision.
The number of arches always coincides with the number of the scallops.  
The mass combinations of total degeneracy are numerically found by Sim\'o (1980). 
We introduced a family of sets of symbol sequences, and 
partitioned the Poincar\'e section by sets of points corresponding to 
sets of symbol sequences. 
We then found that an arch consists of sets of points whose symbol sequences 
are arranged according to a simple rule (see \S 4.1 of Paper I).
The composition of arches changes each time the number of scallops increases, 
that is, the mass parameters pass through a totally degenerate case. 
We observed that a number of germs bifurcate from arches and then 
germs from different arches construct new arches. However, we could not 
understand the dynamics behind the behaviour of germs.
\par
In the present paper, we also study the structure change of the 
Poincar\'e section as the central mass varies (we only consider 
the symmetric mass combinations). However, in contrast to Paper I, 
where the structure of the Poincar\'e section was considered in 
relation to triple collision, here the structure is analysed 
in relation to the motion of periodic orbits bifurcated from the Schubart 
orbit. From another point of view, the azimuthal structure of the 
Poincar\'e section with centre at the Schubart point has been studied 
in Paper I, whereas the radial structure with centre at the Schubart
point will be studied in the present paper.
\par
The rotation number with respect to the (Schubart) fixed 
point will be introduced. 
General tendency of the motion of the bifurcated orbits from the 
fixed point is analysed. The rotation number monotonically decreases 
as $a$ increases. 
As we shall see, there is the sequence of 
rotation numbers such that the corresponding periodic orbits has 
significant influence on the Poincar\'e section. these orbits will be 
called {\it dominant periodic orbits}. We mainly concern with these 
orbits. 
As the mass parameters change, the dominant periodic orbits bifurcate, 
recede from the fixed point, and their stability region shrink.
We will relate these behaviours of periodic points to the growth of 
the germs observed in Paper I.
\par
The organisation of the present paper is follows.
In \S 2, we introduce the equations of motion of our system.
We define the Poincar\'e section and introduce the
{\it rotation number} in the fixed-point-centric coordinates.
The \S 3 is for the results. In \S 3.1, the bifurcation of periodic
points is analysed as the mass parameter changes. 
In \S 3.2, we find that periodic points with special rotation number
are dominant over the structure of the Poincar\'e section. In \S 3.3, 
we show in detail the radial motion of the periodic point with the 
growth of the germs as the mass parameter changes. Finally in \S 4, 
we summarise our results.

\section{Method}
\subsection{Equations of Motion and Initial Points}
  We introduce the equations of motion, the Poincar\'e section, and
parameters describing mass combination of the particles (MH1989). Then,
we introduce symbol sequences recording the collisional history of
orbits (TM2000a).
\par
  Let $m_1,m_0$, and $m_2$ denote the masses of the three particles from
left to right on the line. Let $(q_1,q_2)$ be the mutual distances
$\overline{m_1m_0}$ and $\overline{m_0m_2}$, and $(p_1,p_2)$
the conjugate momenta to $(q_1,q_2)$.
The Hamiltonian $H$ is written, with the kinetic energy $K$ and the 
force function $U$, as
\begin{eqnarray}
  H&=&K-U, \\
  K&=& \frac{1}{2}\Big(
           \frac{1}{m_1}\kern-0.2em + \kern-0.2em \frac{1}{m_0}
       \Big)p_1^2 +
       \frac{1}{2}\Big(
        \frac{1}{m_0}\kern-0.2em+\kern-0.2em\frac{1}{m_2}
       \Big) p_2^2 -
       \frac{p_1p_2}{m_0}, \nonumber \\
  U&=& \frac{m_1m_0}{q_1} + \frac{m_0m_2}{q_2} + \frac{m_1m_2}{q_1+q_2}.
  \label{Hamiltonian}
  \nonumber
\end{eqnarray}
In this paper, we consider the case that total energy $E$, where $H = E = const.$, is negative.
We restrict ourselves to $E=-1$, since negative energy system can be brought to $E=-1$ 
using the homogeneity of $U$. We here adopt variables used in MH1989. For readers' 
convenience, we repeat the formulation.
Let us consider the canonical transformation from $(q_i,p_i,t)$ to
$(Q_i,P_i,t')$ defined by
\begin{eqnarray}
  q_i&=&Q_i^2,\ p_i=2p_iQ_i\ \ (i=1,2) \label{hoge}\\
  dt&=&q_1q_2dt'.
\end{eqnarray}
Then the new Hamiltonian reads
\begin{eqnarray}
  \Gamma&=&q_1q_2(H-E),
\end{eqnarray}
and the equations of motion become
\begin{eqnarray}
  \frac{dQ_i}{dt'}=\frac{\partial\Gamma}{\partial P_i},\quad
  \frac{dP_i}{dt'}=-\frac{\partial\Gamma}{\partial Q_i}\quad(i=1,2).
  \label{regular.eq.5}
\end{eqnarray}
The transformed equations (\ref{regular.eq.5}) show that binary collisions
($Q_1=0,Q_2\ne0$ or $Q_2=0,Q_1\ne0$) are regularised, whereas triple
collision ($Q_1=Q_2=0$) is still singular. The solution to Eq.
(\ref{regular.eq.5}) is continued beyond binary collision.
A binary collision is interpreted as an elastic bounce in the physical
coordinates.
\par
  The equations of motion allow a solution $q_1(t)=\Lambda q_2(t)$
for all $t$ for which the solution is defined.
This is called the {\it homothetic solution} (\cite{IN1972}).
Here $\Lambda$ is a constant depending on masses. This dependence 
is given by the following equation: 
\begin{eqnarray}
\frac{-m_1 + zm_0 + m_2}{m_1 + m_0 + m_2} = 
  \frac{z^5 -2z^3 + 17z}{z^4-10z^2 - 7},& {\rm where} &  \Lambda=\frac{1+z}{1-z}.
\end{eqnarray}

The intersection of the energy hyper-surface $H(q,p)=E$ and the
hyper-surface $q_1=\Lambda q_2$ is a two-dimensional surface.
We take this surface as the Poincar\'e section $\overline{\Pi}$.
In the following, we introduce coordinates $(\theta,R)$ on
$\overline{\Pi}$. The variable $R$ is defined by
\begin{eqnarray}
  R=\frac{1}{2}(q_1+q_2)\quad {\rm where}\quad {q_1=\Lambda q_2}.
  \label{eq7}
\end{eqnarray}
With solving Eq. (\ref{eq7}) to obtain $q_1$ and $q_2$ for a given $R$, 
the value of $K$ is determined as
\begin{equation}
  K = E+U(q_{1},q_{2})\quad {\rm with}\quad 
      q_{1} = \frac{2\Lambda R}{1+\Lambda},\ 
      q_{2} = \frac{2R}{1+\Lambda}. \label{K(R)}
\end{equation}
For this value of $K$, the parameter $\theta$ determines the ratio
of $\dot{q}_1$ to $\dot{q}_2$:
\begin{eqnarray}
  & & \kern-2em
      \sqrt{K}\cos\theta =
      \sqrt{A-\kappa {r_2}^2}\dot{q}_1 - \sqrt{B-\kappa {r_1}^2}\dot{q}_2
      \nonumber \\
  & & \kern-2em
      \sqrt{K}\sin\theta =
      r_2\sqrt{\kappa} \dot{q}_1     - r_1\sqrt{\kappa} \dot{q}_2,
      \label{No12}
\end{eqnarray}
where
\begin{eqnarray*}
  & & A=\frac{m_1(m_0+m_2)}{2M},\
      B=\frac{m_2(m_0+m_1)}{2M},\
      C=\frac{m_2m_0}{2M},\
      M=m_1+m_0+m_2,\\
  & & r_1=\frac{2\Lambda}{\Lambda+1},\  r_2=\frac{2}{\Lambda +1},\
     \kappa = \frac{4AB-C^2}{4(A{r_1}^2+B{r_2}^2+Cr_1r_2)},\ 
     0\leq\theta < 2\pi.
\end{eqnarray*}

The variable $R$ takes its maximum value $R_{\rm max}$ when $K=0$.
Substituting $K=0$ in Eq. (\ref{K(R)}), we obtain 
\begin{eqnarray}
  R_{\rm max}=\frac{|E|}{2}\Big[
     \frac{\Lambda+1}{\Lambda}m_0m_1+(\Lambda+1)m_0m_2+m_1m_2 \Big].
\end{eqnarray}
We denote by $\Pi$ the side of the Poincar\'e section with 
$0\leq\theta < \pi$ and by $\Pi^*$ the other side with 
$\pi\leq\theta < 2\pi$. By the definition of $\theta$ 
in Eq. (\ref{No12}), $\dot{\VEC q}(\theta+\pi)=-\dot{\VEC q}(\theta)$, 
where we denote by $\dot{\VEC q}(\theta)$ the $(\dot q_1,\dot q_2)$ 
for given $\theta$.  Since the original equations of motion are 
invariant under the transformation $(\dot{q}_i,t) \to (-\dot{q}_i, -t)$, 
it is enough to integrate orbits toward the future for each point 
$(\theta, R)$ in $\Pi \cup \Pi^*$.  Nonetheless, we only study the structure 
of the surface $\Pi$ for future integration. The reason why we do
so is that an orbit intersects alternately with $\Pi$ and $\Pi^*$ 
until the corresponding triple system disintegrates into a binary and
a single particle. Only certain orbits starting from the immediate
escape region intersect with only $\Pi$ 
or $\Pi^*$. Since we are interested in the structure of chaotic scattering 
region, such escape orbits can be neglected.
\par

We assume the total mass of the particles to be three without loss
of generality and introduce parameters $a$ and $b$ to represent the masses:
\begin{eqnarray}
  & & \kern-3em m_1=1-a-b,\ m_0=1+2a,\ m_2=1-a+b, \label{No21} \\
  & & {\rm where}\ a>-1/2,\ b>0,\ a + b < 1. \nonumber
\end{eqnarray}
The parameters move in a triangular area of the $(a,b)$-plane.
We will call this a {\it mass triangle}. In Paper I, we saw the 
structure change of the Poincar\'e section over the mass triangle. 
In the present paper, we only consider the symmetric mass 
configurations: $b=0$.  
The reason is that we saw in Paper I that asymmetry of the mass 
configuration in many cases does not add topologically new structure
to the symmetric case.

\par
  There are three types of collision in our system: left-centre
($q_1=0,q_2\ne0$), centre-right ($q_2=0,q_1\ne0$), and triple
($q_1=q_2=0$) collisions. Let us denote these collisions by
symbols `1', `2', and `0', respectively. 
An orbit of the partly regularised equations Eq. (\ref{regular.eq.5}) 
repeats binary collisions until a triple collision takes place. 
These collisions
can be recorded as a sequence of the symbols. We study the
evolution of orbits using the symbol sequences instead of orbits
themselves. It is to be noted that we encode into symbol sequences
the future behaviour of the orbits starting at points of the Poincar\'e 
section. So symbol sequences are not bi-infinite but singly infinite
to the future.

\par
%
Integration of orbits are carried out for grid points. 
We introduce the grid points $(\theta_i,R_i)$ with different resolution on $\Pi$:
\begin{eqnarray}
& & (\theta_i,R_i)=(\pi\times\frac{i}{\theta_{\rm div}},
   R_{\rm max}\times \frac{j}{R_{\rm div}}),
   \hbox{ where }i=0,\cdots,\theta_{\rm div}-1,\ j=1,\cdots,R_{\rm div}, \nonumber\\
& & \hskip4em\raise1.5ex\hbox{and }
        \left\{\begin{array}{ll}
           {\rm (i)}& \theta_{\rm div}=540,\ R_{\rm div}=300     \\ 
           {\rm (ii)}& \theta_{\rm div}=1800,\ R_{\rm div}=1000  
        \end{array}\right.
\end{eqnarray}
Most of structures on the Poincar\'e section and their change due to 
the mass variation can be resolved with resolution (i). 
Detailed structures may be lost, if one uses resolution (i). 
However, once we understand  certain fine structures with resolution (ii), 
we can use resolution (i) to follow global behaviour of the structure 
of the Poincar\'e section when mass parameters are changed. 
We have obtained the first 64 digits of symbol
sequences for the grid points through integration of orbits.
These constitute the basic data in the rest of the paper for the
distribution of symbol sequences on $\Pi$. 
We integrate Eq. (\ref{regular.eq.5}) using DIFSY1, which is an 
implementation (\cite{PT1999}) of Bulirsch-Stoer method. 
  In our preliminary research (\cite{ST2004}), we introduced
cylinders (sets of symbol sequences with given words where a word is a
finite sequence of symbols) $S_{c,j}$, $S_{c,\infty}$ and $S_{c}$:
\begin{eqnarray}
& &
  S_{c,j} \equiv \left\{
    \begin{array}{l}
        { \{(21)^i(2)^j\cdots|i\geq 0,j\geq 1\}}\quad {\rm for}\ {c=2i+1}\\
        { \{(21)^i(1)^j\cdots|i\geq 1,j\geq 1\}}\quad {\rm for}\ {c=2i}
    \end{array} \right. \label{def.Sc.1}\\
& &
  S_{c,\infty} \equiv \left\{
    \begin{array}{l}
        { \{(21)^i(2)^{\infty}|i\geq 0} \} \quad {\rm for}\ {c=2i+1}\\
        { \{(21)^i(1)^{\infty}|i\geq 1} \} \quad {\rm for}\ {c=2i}
    \end{array} \right. \label{def.Sc.2} \\
& &
  S_c \equiv \cup_{j<\infty} S_{c,j}.
  \label{def.Sc.3}
\end{eqnarray}
Let $\reg{c}$ and $\reg{(c,\infty)}$ denote the regions on $\Pi$ whose
symbol sequences belong to $S_c$ and $S_{c,\infty}$.
The three basic regions in HM1993 correspond to our regions
as follows:
\begin{center}
  \begin{tabular}{lll}
    \hline
      Immediate escape region & $\reg{(c,\infty)}$  & $1\leq c \leq \nroot$
\\
      Chaotic scattering region & $\reg{c}\cup\reg{(c',\infty)}$ &
         $1 \leq c < \infty, \ \nroot+1 \leq c' < \infty $\\
      Schubart region & $\reg{c}$ & $c=\infty$ \\
    \hline
  \end{tabular}.
\end{center}
\begin{figure}
\begin{center}
\includegraphics[width=100mm,height=62.5mm]{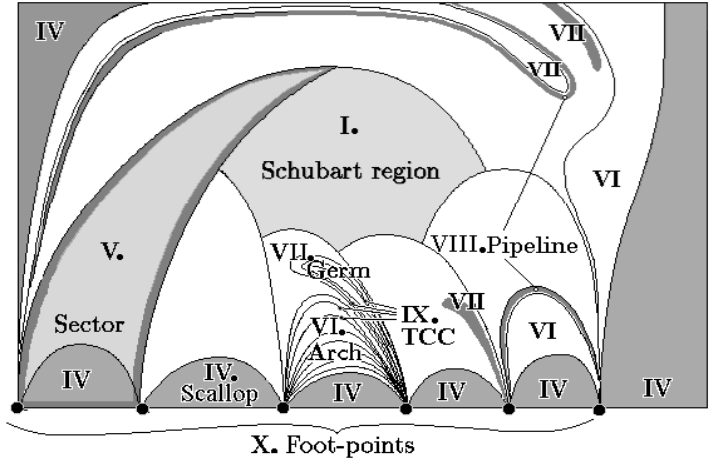}
\end{center}
\caption{
The schematic illustration of the structure on the Poincar\'e
section. The Poincar\'e section consists of the Schubart region (I) and 
sectors (V). A sector (V) includes one scallop (IV), one arch (VI), 
and optional germs (VII) and pipelines (VIII). 
In the third sector from the left, it is shown that both an arch and 
a germ consist of the strata of triple collision curves. In the blank 
area of the third sector, there are other germs 
and/or pipelines, although they are not drawn. 
The pipeline (VIII) is an object which connects two germs in different 
sectors. Note that there are narrow escape regions also inside 
the sectors.
}
\label{struct00}
\end{figure}
\par
For later convenience, we recall the structure of the Poincar\'e section 
and names for its elements obtained in Paper I (Fig. \ref{struct00}). 
The separatrices running from the vertices of 
(I) the {\it Schubart region} divide the outside of the Schubart region 
into (V) {\it sectors}. A sector includes (IV) one {\it scallop}. In 
the remainder of the sector, (IX) triple collision curves are foliated. 
This foliation makes one or more blocks. One block is  
arch-shaped (VI, {\it arch}) and the other blocks are germ-shaped 
(VII, {\it germ}). A germ bifurcates from an arch, as $a$ 
changes. Germs that bifurcate from different arches finally gather to 
construct a new arch and to reconstruct other arches. 
As is shown later, the growing process of germs is related 
to the movement of the periodic orbits bifurcated from the Schubart orbit.

\subsection{Poincar\'e map and rotation number}
  We here define the Poincar\'e map on the Poincar\'e section $(\theta,R)$. 
An orbit starting from the Poincar\'e section repeats the intersection 
with $\Pi$ and $\Pi^*$ alternately. 
When an orbit intersects with $\Pi$ (resp. $\Pi^*$) at $(\theta,R)$
and $\Pi$ (resp. $\Pi^*$) again at $(\theta'',R'')$, we define a map T
from $(\theta,R)$ to $(\theta'',R'')$.
\begin{equation}
  T:(\theta,R)\mapsto(\theta'',R''), \hbox{ where }
    (\theta,R),(\theta'',R'')\in\Pi\hbox{ or }
    (\theta,R),(\theta'',R'')\in\Pi^*
\end{equation}
\par
  Equation (\ref{regular.eq.5}) has a simple periodic solution, the 
so-called Schubart orbit (Schubart, 1956), whose symbol sequence is
$(21)^\infty$. The intersection $\VEC P_0 = (\theta_0,R_0)$ of the
Schubart orbit with $\Pi$ is a fixed point of the map, namely 
$\VEC P_0=T(\VEC P_0)$.  The linear stability of $T$ around $\VEC P_0$ 
over the mass triangle is studied by HM1993. Approximately 
in the mass combinations such that $m_1<m_0<m_2$, $\VEC P_0$ 
is hyperbolic whereas in the other combinations it is elliptic. 
The stability region is called the Schubart region. 
A point $\VEC P$ such that $\VEC P = T^q(\VEC P)$ is 
called a {\it q-periodic point} and the sequence 
$\{\VEC P,T^1(\VEC P),\cdots,T^{q-1}(\VEC P)\}$ is called a {\it q-periodic orbit}.
We are interested in the periodic orbits that bifurcate from the fixed point
as the mass parameter $a$ is changed. 
\par
\def\gTr{{\rm g}_\theta} 
\def\gR{{\rm g}_R} 
In order to describe the elliptic motion around $\VEC P_0$, we introduce the
{\it rotation number}. The rotation number is the averaged number of rotations 
per iterate of $T$.
First, we introduce polar coordinates $(D,A)$ with centre at 
$\VEC{P}_0=(\theta_0,R_0)$:
\begin{eqnarray*}
& & D\cos A = (\theta - \theta_0) \gTr,\quad 
    D\sin A = (R-R_0) \gR, \\
\end{eqnarray*}
where $\quad \gTr = 180/\pi$ and $\quad \gR=100/R_{\rm max}$.
The values of $\gTr$ and $\gR$ are arbitrarily chosen such that
both $\theta$ and $R$ equally contribute to the values of $D$ and $A$.
Now the Poincar\'e map is expressed as $T: (D,A) \mapsto (D',A')$.
We then introduce the effective rotation number $\nu_n(D,A;a)$ and  
the (exact) rotation number $\nu_\infty(D,A;a)$ at $(D,A;a)$ in the 
following equations: 
\begin{eqnarray}
  & &{\rm diff}(A',A) = \bigg\{ 
     \begin{array}{ll}
       A'-A         & ({\rm if}\ A'-A\geq 0) \\ 
       A'-A + 2\pi  & ({\rm if}\ A'-A < 0) 
     \end{array} \label{diff_A} \\
  & &\nu_n((D,A);a) = \frac{1}{2n\pi}\sum_{i=1}^n 
         {\rm diff}(A^{(i)},A^{(i-1)}),\quad 
     {\rm where}\ (D^{(i)},A^{(i)})=T^i(D,A) \label{rot_num_com} \\
  & & \nu_\infty((D,A);a) = \lim_{k\to\infty} \nu_k(D,A;a) 
   \kern2em \hbox{(if the limit exists)}. 
  \label{rot_num}
\end{eqnarray}
We here remark several points related to the numerical calculation of 
the rotation number. The first point is the selection of $n$ of the 
$\nu_n((D,A);a)$. In our approach, we first take an fixed integer $N$ and 
iterate the mapping $N$ times. We then find a number $n$ ($1\leq n \leq N$) 
such that ${\rm diff}(A^{(n)},A) = \min _{1\leq i \leq N} \{{\rm diff}(A^{(i)},A)\}$.
The second point concerns the rotation number of periodic points. 
Suppose $p$ and $q$ are integers. If $\VEC{P}$ is a periodic point, 
then there exist integers $p$ and $q$ such that 
$\nu_q(\VEC{P};a)=\nu_\infty(\VEC{P};a)=p/q$ (by the above 
method of selection of $n$, $n=q$).
This coincidence of the effective and exact rotation numbers helps us
to find the periodic points: the points on the 
contour with $\nu_q(\VEC{P};a)=p/q$ on the the Poincar\'e section 
are the candidates of the periodic points with $\nu_\infty(\VEC{P};a)$.
The third point is that the rotation number $\nu_\infty(\VEC{P}_0;a)$ 
is calculated from the eigenvalues of 
linearised map of $T$ at $P_0$. When this value is rational, 
the fixed point and the periodic points are degenerated. 
If the mass parameter $a$ is changed, the periodic points bifurcate
from fixed point $\VEC P_0$.

\subsection{Search for the periodic points}
\begin{figure}
\leavevmode
\hfil
\vbox{\hsize=75mm
\includegraphics[width=75mm,height=52mm]{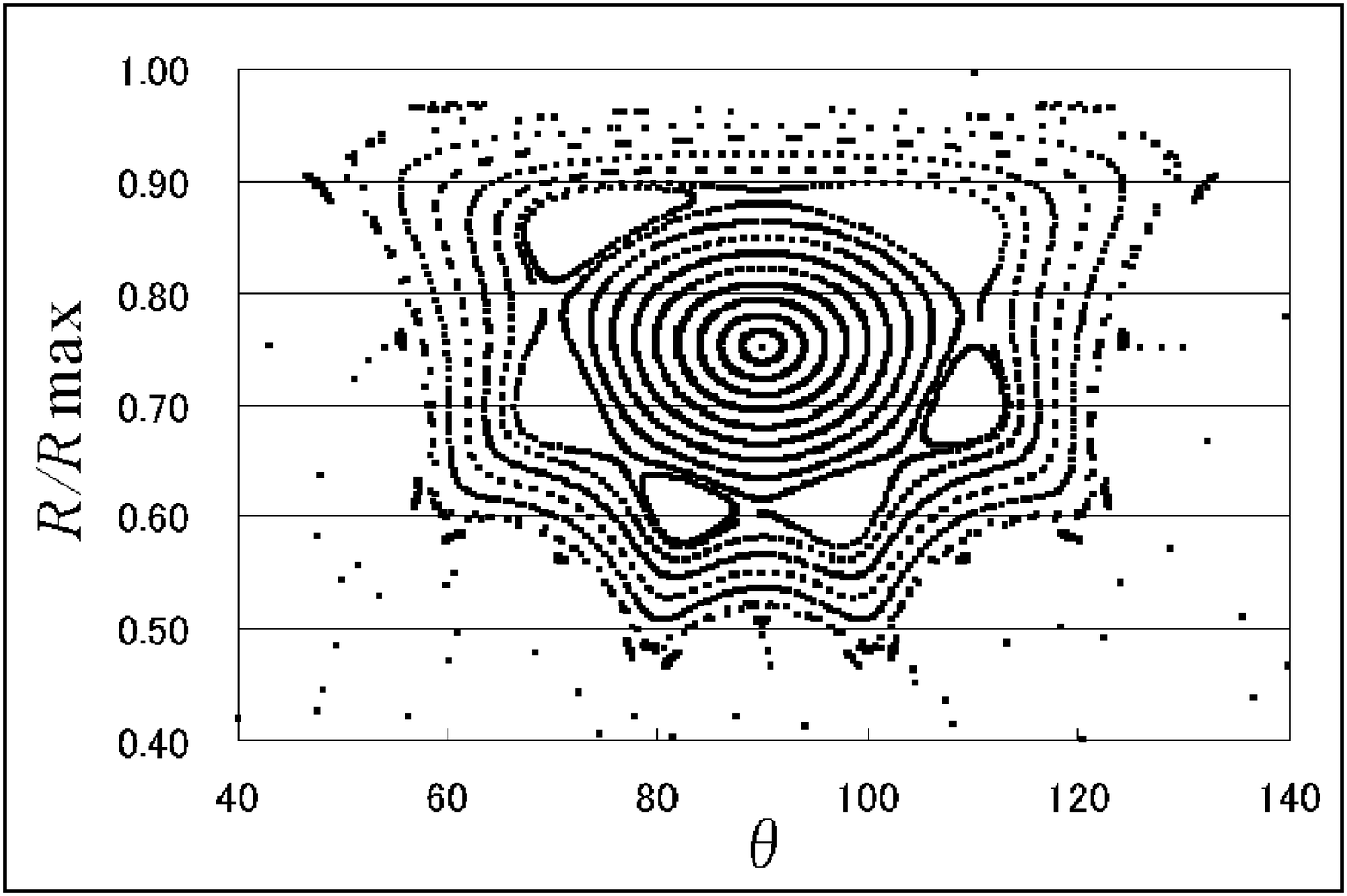}
}
\caption{
A Poincar\'e map in order to select candidates for periodic points. 
In this map, $a=-0.25$ and the expected periodic points are those with 
$\nu=4/6$.  The centres of elliptic motions under the Poincar\'e 
map are the candidates. 
}
\label{find_candidate}
\end{figure}

  We explain how to detect the periodic points. First, we have to 
calculate $\lim_{D\to 0} \nu_\infty((D,A);a)$ for various values of $a$. 
For a given rotation number $p/q$, we can determine the mass parameter $a$ 
where corresponding periodic points appear: solve 
$\nu_\infty(\VEC P_0;a)=p/q$ for $a$. Second, we find candidates at 
a mass parameter a little distant from the exact bifurcation. 
The candidates for periodic points are found from the Poincar\'e 
map such as is shown in Fig. \ref{find_candidate}. The candidates for elliptic points 
can be taken at the centre of the libration and for hyperbolic points 
at the saddle between two librations. 
%
%
Finally, we obtain a periodic point through the Newton-Raphson 
method Eq. (\ref{NRM}) starting from an arbitrary point on the contour.
This method is used in MH1991 to find the fixed point.
\begin{equation}
  \vec{X}_{n+1}=\vec{X}_{n} - 
  \Bigg[\del{T(\vec{X}_n)}{\vec{X}_{n}}-1\Bigg]^{-1}
  (T^p(\vec{X}_{n})-\vec{X}_{n})=0.
  \label{NRM}
\end{equation}

\section{Results}
We now show the behaviour and influence of the periodic points on the 
Poincar\'e section. First, their bifurcation and radial movement away 
from the Schubart orbit will be discussed. The values of parameter $a$ 
where the periodic points bifurcate can be determined from the rotation 
number, $\nu_\infty({\VEC P_0}; a)$, of the Schubart orbit as a function 
of $a$. The relation between radial movement and the variation of $a$ 
will be studied.
Second, the number of orbits which appear in respective bifurcations 
for special rotation numbers will be discussed. Finally, the influence of 
such special periodic orbits on the Poincar\'e section will be discussed.


\subsection{Bifurcation as the parameter changes}
\begin{figure}[htb]
\begin{center}
  \includegraphics[width=5.16in,height=3.77in]{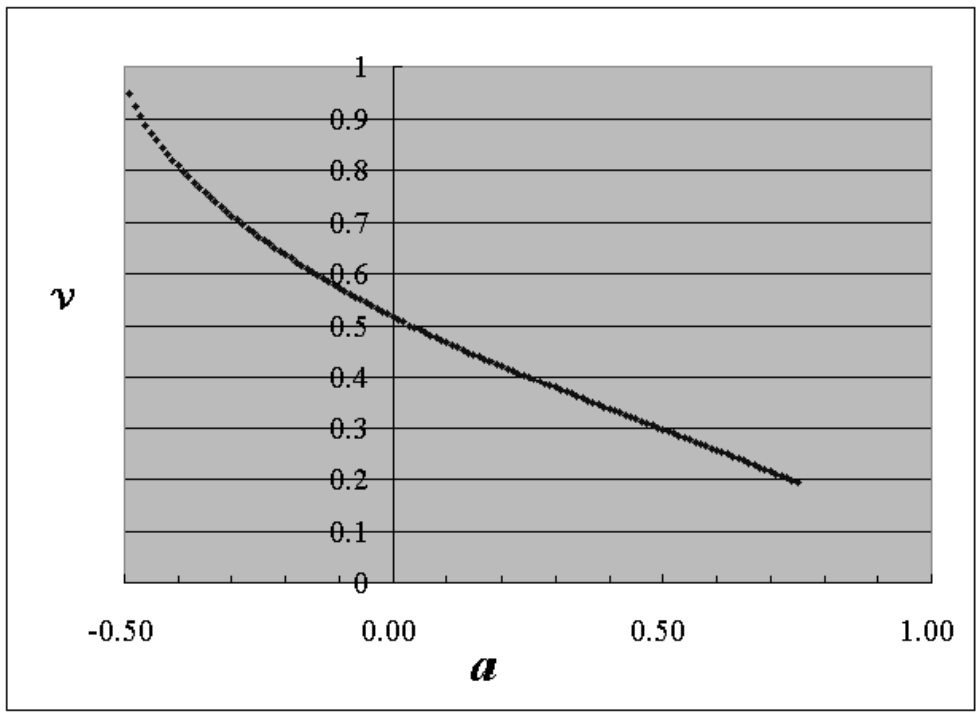} \par
  \caption{The rotation number $\nu(\VEC P_0,a)$ as a function of $a$.}
  \label{alpha_a_curve}
\end{center}
\end{figure}

\begin{figure}[htb]
\leavevmode
\vbox{\hsize=75mm
(a) The distance from $\VEC P_0$ \\
\includegraphics[width=75mm,height=48.6mm]{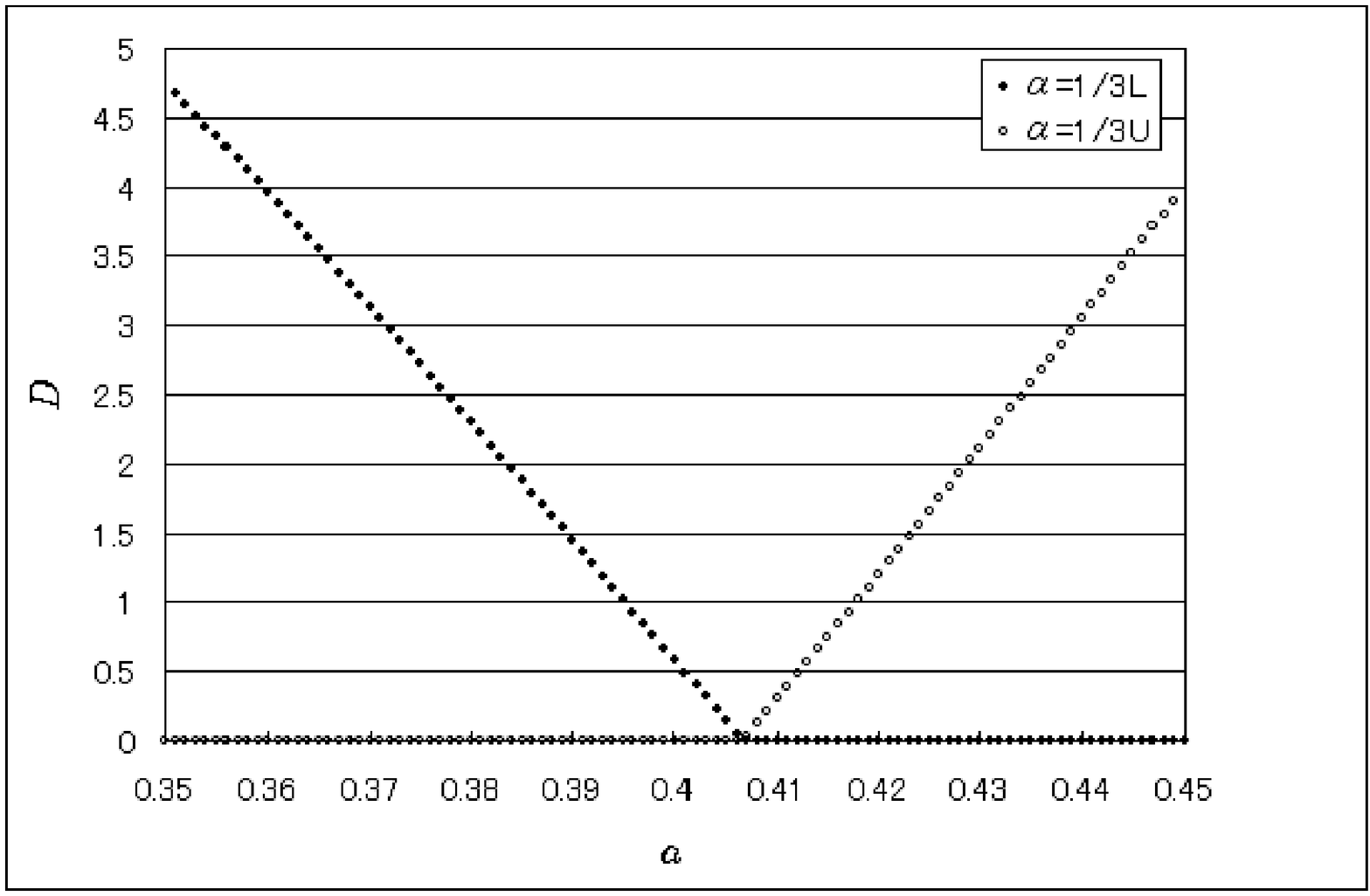}\par
}
\hskip3mm
\vbox{\hsize=77mm
(b) The inversion of triangle's orientation \\
\leavevmode
\includegraphics[width=37mm,height=48.6mm]{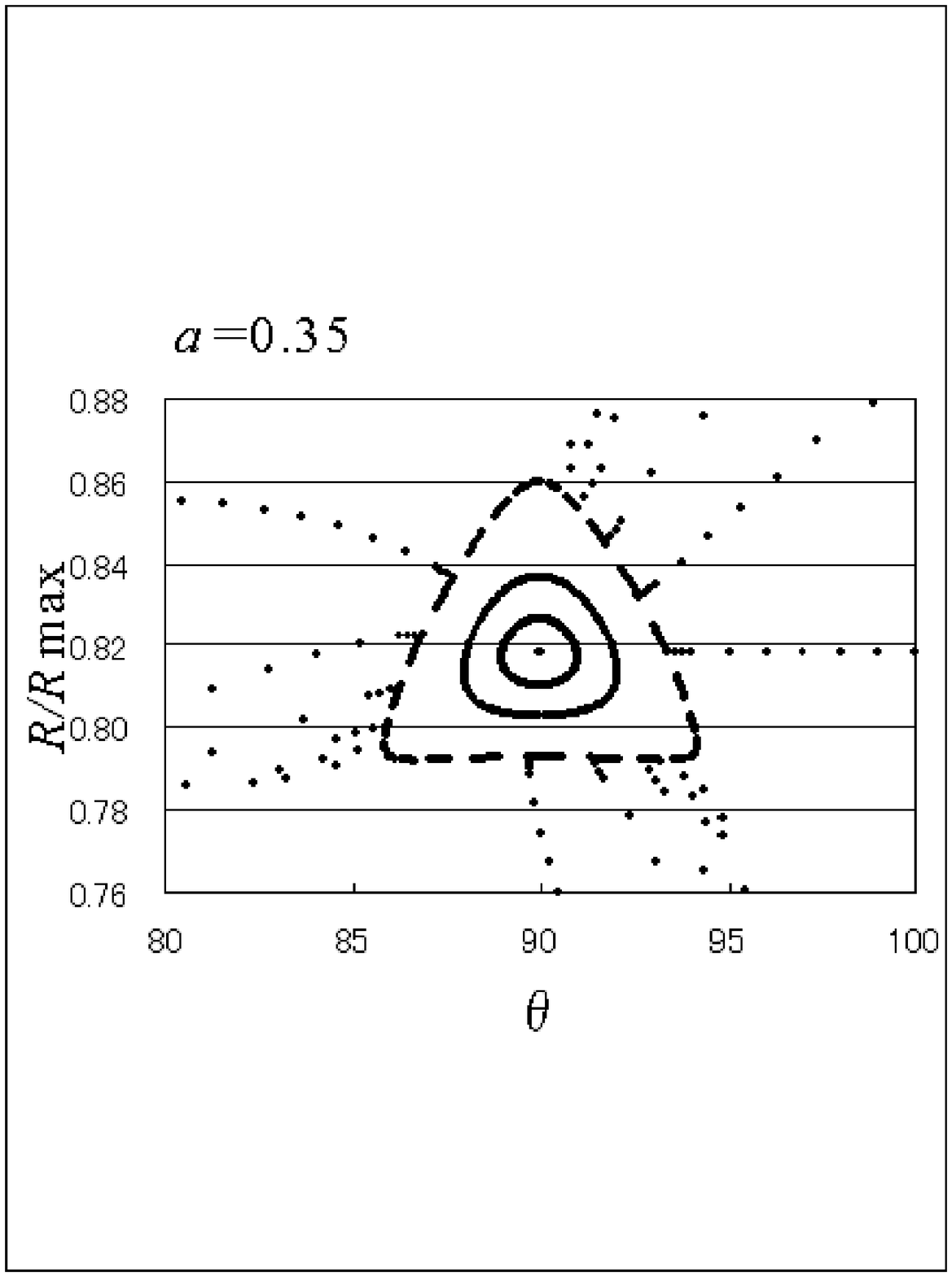}
\includegraphics[width=37mm,height=48.6mm]{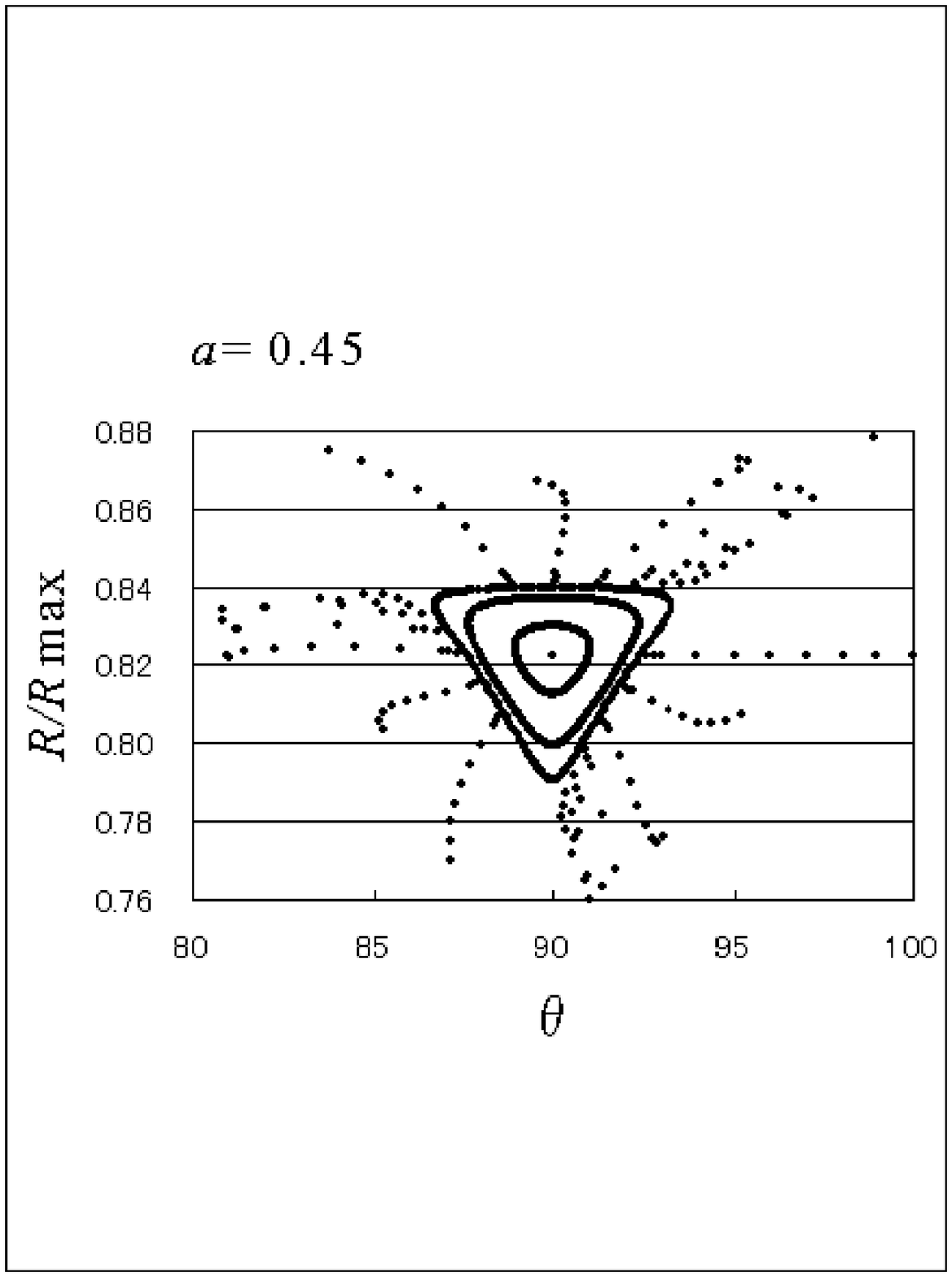}
}
  \caption{The radial movement of the periodic points with 
  $\nu_\infty=1/3$ from $\VEC P_0$. (a) the distance between one of 
  these point and $\VEC P_0$. 
  }
  \label{Da_pp1_3}
\end{figure}

  In this subsection, we show how periodic points bifurcate and move
outward from the fixed point $\VEC P_0$, when the parameter $a$ is 
changed.
As is stated in section 2.2, period-$q$ points bifurcate if 
$\nu_\infty(\VEC P_0; a)=p/q$ with integers $p$ and $q$. 
In order to know when the periodic orbits bifurcate, we calculate 
$\nu_\infty(\VEC P_0; a)$ as a function of $a$ (Fig. \ref{alpha_a_curve}). 
For several rotation numbers, we show the values of $a$ in Table 
\ref{a_bifurcate}. Figure \ref{alpha_a_curve} shows $\nu_\infty(\VEC P_0; a)$ 
is monotonic:
\begin{equation}
  \del{\nu_\infty(\VEC P_0; a)}{a} < 0. \label{monotonicity}
\end{equation}
Therefore for each rational number $p/q$, the periodic points with 
$\nu_\infty=p/q$ bifurcate once and for all. 
\begin{table}[htb]
\begin{center}
\begin{tabular}{ccccccc}
\hline \hline 
$\nu_\infty$ & 1/4  & 1/3  & 2/4    & 3/5   & 4/6   & 3/7   \\ 
$a$          & 0.61 & 0.41 & -0.019 & -0.15 & -0.25 & -0.31 \\ 
\hline \hline
\end{tabular}
\caption{The values of $a$ at which the periodic points bifurcate 
for several rotation numbers}
\label{a_bifurcate}
\end{center}
\end{table}
%
%
Note that this graph shows that the period-3 orbit bifurcates at 
$a=a_{1/3}\fallingdotseq0.41$. According to Moser (1958), generally 
in Hamiltonian systems when the period-3 orbit bifurcates, 
the mother periodic orbit becomes unstable. As an application of 
this result, he tried to explain the origin of the Kirkwood gaps. 
Also in our system, 
the instability at the bifurcation of the period-3 orbit was
confirmed by Hietarinta and Mikkola (1993). 
We have followed the position of the periodic 
points with $\nu_\infty = 1/3$ on the Poincar\'e section with 
continuously changing $a$.  As a result, when $a$ is decreased, 
with a passage $a=a_{1/3}$, the triangle formed by these periodic points 
approaches ($a>a_{1/3}$), collides ($a=a_{1/3}$) with, and recedes  
($a<a_{1/3})$ from $\VEC P_0$ (Fig. \ref{Da_pp1_3}(a)). The orientation of 
the triangle is inverted before and after $a_{1/3}$ 
(Fig. \ref{Da_pp1_3}(b)).  Similar behaviour was observed by 
H\'enon (1970) in Hill's case of the restricted three-body problem. 
In his case, mother periodic orbit is the retrograde satellite orbit. 
\par 
  We want to know the direction of change of $a$ for the bifurcated 
periodic points to move outward from $\VEC P_0$.
Suppose that the periodic points bifurcate at $a=a_{p/q}$ and exist 
at $a=a_{p/q}+da$ as the points with a finite distance $dD$ from 
$\VEC P_0$. The sign of $da$ can be obtained from  
Eqs.(\ref{monotonicity}) and (\ref{radial_derivative}), as follows. 
The total differential of $\nu$ as a function of $a$ and $D$ at 
$\VEC P_0$ is 
$$
  d\nu = \Big(\del{\nu}{a}\Big)\Big|_{\VEC P_0}da + \Big(\del{\nu}{D}\Big)\Big|_{\VEC P_0}dD.
$$
Since $d\nu=0$ on periodic points, we obtain
\begin{equation}
  da = - \Big(\del{\nu}{D}\Big/\del{\nu}{a}\Big)\Big|_{\VEC P_0} dD.
  \label{da}
\end{equation}
\par
\begin{figure}[htbp]
\begin{center}
\leavevmode
\vbox {\hsize=75mm
\includegraphics[width=75mm,height=52mm]{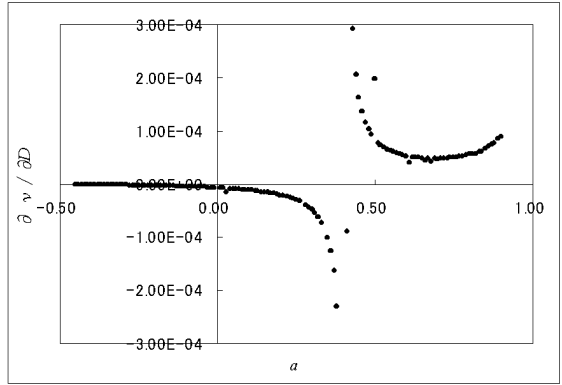}
}
\end{center}
\caption{Dependence of $\d\nu(D,a)/\d D$ on $a$. The sign of $\d\nu(D,a)/\d D$
changes at $a=a_{1/3}$.}
\label{alpha_D_curve}
\end{figure}

  We only need the sign of $\d \nu_\infty(\VEC P_0;a)/\d D$, since we already 
know the sign of $\d \nu_\infty(\VEC P_0;a)/\d a$. 
We approximate the derivative $\d \nu_n(\VEC P;a)/\d D$ by the difference
$(\nu_n((D+\Delta D,A);a)-\nu_n((D,A);a))/\Delta D$ for small $\Delta D$. 
Since $\d \nu_n(\VEC P_0;a)/\d D$ should be independent to $A$, 
we select arbitrarily the value of $A$: $A=0$. 
Moreover, we have $\nu_\infty(\VEC P_0;a) \fallingdotseq \nu_n(\VEC P_0;a)$
for large $n \leq N$: $N=256$. Finally we put $\Delta D = 0.1$. 
Figure \ref{alpha_D_curve} 
shows $\d \nu_\infty(\VEC P_0;a)/\d D$ as a function of 
mass parameter $a$. The figure shows that the sign of 
$\d \nu_\infty(\VEC P_0;a)/\d D$ is 
\begin{equation}
   \left\{
   \begin{array}{cc}
      \displaystyle \del{\nu_\infty(\VEC P_0;a)}{D} > 0 & (\hbox{for }a>a_{1/3})\\
      \raise4ex\hbox{}
      \displaystyle \del{\nu_\infty(\VEC P_0;a)}{D} < 0 & (\hbox{for }a<a_{1/3})\\
   \end{array}
   \right.
   \label{radial_derivative}
\end{equation}
\par
Taking into account Eqs.(\ref{monotonicity}) and (\ref{radial_derivative}), 
and $dD>0$, we find that $da$ is positive for $a>a_{1/3}$ 
and negative for $a<a_{1/3}$. The recession from $a=a_{1/3}$, 
irrespective of the sign of $a$, yields successive recessions of 
the periodic points from the Schubart orbit. This implies that absense of the 
periodic points at $a=a_{1/3}$ (except for the periodic points 
independent of $\VEC P_0$). 

\subsection{Dominant Periodic Orbits}
  In the present subsection, we will find periodic orbits and follow their 
motions for various rotation number. These results show us that the periodic 
orbits with rotation number $(n-2)/n$ dominate the structure of 
the Poincar\'e section. We call these the {\it dominant periodic 
orbits}. We discuss a few features of the dominant periodic orbits.
On the other hand, periodic orbits with the other rotation 
numbers have too small stable regions to numerically find the location 
on the Poincar\'e section, so we do not consider these.
\par 
  According to our numerical results, there is a rule for the 
number of dominant periodic orbits. 
Suppose that the rotation number of these orbits is 
$\nu_\infty=(n-2)/n$. Indeed, for even $n$, the period is not
$n$ but $n/2$, and therefore, if we would strictly write the 
rotation number, it should be $\nu_\infty=(n/2-1)/(n/2)$. 
We have found the following rule for the number of periodic points.

\begin{figure}[h]
  \leavevmode
  \vbox {\hsize = 80mm 
  (a) $\nu=3/5$ \par
  \hfil\includegraphics[width=60mm, height=40mm]{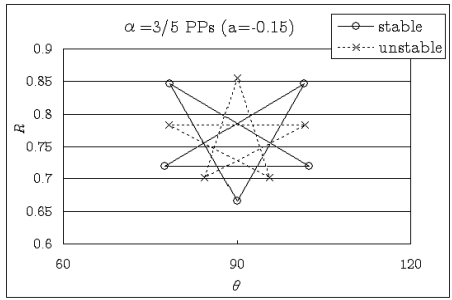} \par
  }
  \vbox {\hsize = 80mm
  (b) $\nu=4/6$ \par
  \hfil\includegraphics[width=60mm, height=40mm]{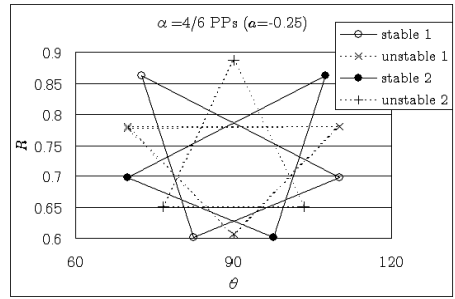} \par
  }
  \caption{Periodic orbits for $\nu_\infty = (n-2)/n$ with 
  (a) $n=5$ and (b) $n=6$. These show the numbers of orbits are 
  2 for odd $n$ and 4 for even $n$. We connect the periodic points which 
  belong to the same orbit. 
  }
  \label{pp_3_5}
\end{figure}

\paragraph{A Rule} 
{\it For the number of dominant periodic orbits with $\nu_\infty=(n-2)/n$, 
the following rule is confirmed from $n=4$ to $n=19$.}
\begin{itemize}
\item {\it the number of orbits is two for odd $n$.} 
\item {\it the number of orbits is four for even $n$. }
\end{itemize}
\noindent
We show an example of this rule in Fig. \ref{pp_3_5} ($n=5$ and 6) 
and Fig. \ref{pp_large_n} ($n=17$ and 18).
For the case of $n=3$ ($\nu_\infty=1/3$), only one unstable orbit appears 
through bifurcation as already seen in Fig. \ref{Da_pp1_3}(b).
It is important to note that according to the 
Poincar\'e-Birkhoff theorem (\cite{Birkhoff1913}) the number of the periodic orbits is 
$2\ell$ with an integer $\ell$, and in the case of the standard map 
(\cite{Chirikov1979}) $\ell$ is believed to be one.
We do not know the number of non-dominant orbits.

\global\def\Z#1#2#3#4{
\vbox { \hsize = 80mm
  (#1) $\nu=#2$, $(a,b)=(#4,0)$\par
  \hfil{\includegraphics[width=80mm, height=80mm]{\IMG/#3}}
}
}

\begin{figure}[h]
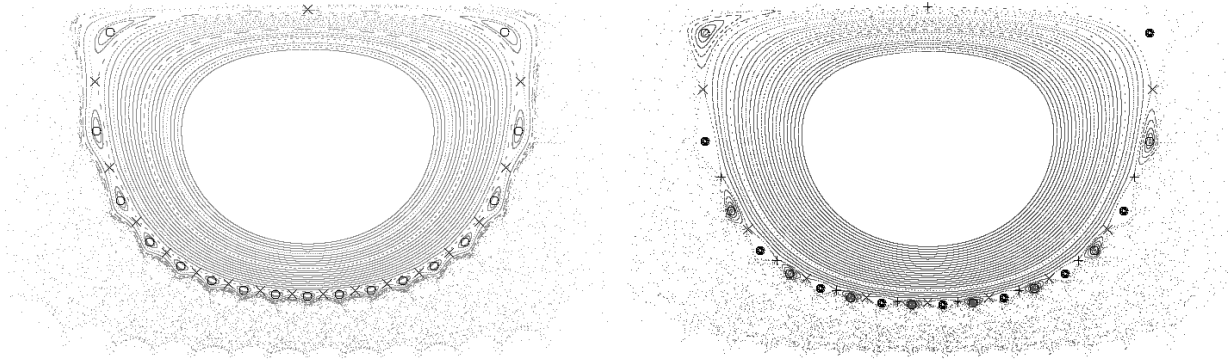

  \leavevmode
  \Z{c}{15/17}{a-0461844_map.\EXT}{-0.461844}
  \Z{d}{16/18}{a-0466010_map.\EXT}{-0.466010}
\caption{Periodic orbits for $\nu_\infty=(n-2)/n$ with (a) $n=17$ and 
$n=18$. For each orbit, its orbital points are plotted by different 
symbols: single and double circles (stable orbits), multiplicative and 
additive symbols (unstable orbits). Two symbols appear for odd case 
($n=17$), whereas four symbols do for even case ($n=18$). The 
background dots shows the Poincar\'e map of segments of several lines 
running from $\VEC P_0$.}
\label{pp_large_n}
\end{figure}
\par

\subsection{Periodic orbits and the structure on the Poincar\'e section}
In the present subsection, we study how the dominant periodic orbits
are related to the structures on the Poincar\'e section.  

Because of the reversal of the bifurcation at $a=a_{1/3}$, we observe 
different movements of periodic orbits for two cases: $da>0$ for 
$a>a_{1/3}$, and $da<0$ for $a<a_{1/3}$. In the latter case, there 
is a nearly periodic change of
the Poincar\'e section, and one cycle is from the bifurcation with 
$\nu_\infty=(n-2)/n$ to that with $\nu_\infty=\{(n+1)-2\}/(n+1)$. 
The change in the former case is basically similar to the one cycle in 
the latter case (Note that for $a>a_{1/3}$ only $\nu_\infty=1/3$ 
has the form $\nu_\infty=(n-2)/n$).  Hence, we propose the scenario for 
the latter case only. The following scenario is based on the observation 
of the cycles starting respectively at $\nu_\infty=2/4, 3/5$, 4/6, 5/7, and 6/8. 
\par
We explain our scenario referring to Fig. \ref{fig280} which corresponds to
the cycle starting at $\nu_\infty=3/5$.
The bifurcated periodic points with $\nu_\infty=(n-2)/n$ ($n \geq 4$) 
recede from $\VEC P_0$ as $a$ decreases.  As the distances become large, 
the stable regions around the periodic points, called {\it islands}, 
influence the shape of the Schubart region. In both Fig. \ref{fig280} 
(a) and (b), five triangular islands can be seen. 
The separatrices which connect the unstable periodic points (and 
envelope the stability region of the stable ones) is of polygramic 
shape. The germs (see Section 2.1) grow along the separatrix. That 
makes the Schubart region be of polygramic shape. When the periodic 
points together with their islands get out of the Schubart region,  
the Schubart region returns to the polygonal shape. The germs now intrude 
between the Schubart region and islands.
As $a$ further decreases, the stable periodic points sink into the gaps 
between arches, and their islands shrink. The germs 
grow and follow the sinking stable periodic points. As a result, 
a set of germs pile up over an arch. The piled germs become a new arch. 
On the contrary, the unstable periodic points stay around the 
vertices of the Schubart region: their separatrix approximates 
the boundary of the Schubart region. If we continue to decrease $a$, 
the next dominant periodic orbits with $\nu_\infty=\{(n+1)-2\}/(n+1)$ 
bifurcate. The evolution of the structure of the Poincar\'e section 
following this bifurcation is similar to the case $\nu=(n-2)/n$. 
It is to be note that there seems to be a correspondence between 
the structure around the Schubart orbit (the fixed point) 
and a structure around a 
bifurcated periodic point. Around the Schubart orbit, there are the 
Schubart region and a number of arches, while around a bifurcated 
periodic point there are the islands and the region filled with germs.
\par
In what follows, let us see numerical evidence that supports our 
scenario. For this purpose, we follow the change of the Poincar\'e 
section over the range of $a$ corresponding to $\nu_\infty=3/5$ and 
4/6. In Fig. \ref{fig280}, the periodic points on the Poincar\'e 
section for the range $-0.150\geq a\geq-0.250$ are displayed. 
The structure of the Poincar\'e section is represented in two 
ways: the triple collision curve and the partitioning according to
the cylinders defined by Eq. (\ref{def.Sc.3}). The separatrices of the 
periodic points are also plotted. 
\par
The radial recession of the periodic 
points and the enlargement of the islands, as $a$ decreases, can be 
seen in Fig. \ref{fig280} (a) $a=-0.150$ and (b) $a=-0.160$.
The germs grow along the separatrix which encloses the islands at
(b) $a=-0.160$, whereas the germs split the Schubart region and the
islands at (c) $a=-0.166$.  This means that the periodic points get 
out of the Schubart region at $a'$ such that $-0.166\leq a'\leq-0.160$. 
After getting out of the Schubart region, islands become smaller. 
This is seen from the transition from (c) $a=-0.166$ to (d) $a=-0.170$. 

The comparison of (d) $a=-0.170$ and (e) $a=-0.186$ shows 
us that the germs grow and follow the stable periodic orbit which is 
sinking between arches, and this growth of the germs results in 
the reconstruction of arches. 
For example, the second arch from the left has reg(2), reg(6), and 
other small regions at (b) $a=-0.160$, whereas reg(2), reg(7), reg(12), 
and other regions at (e) $a=-0.186$. Note that the head of the 
germ $\reg{5,\cdots}$ (the upper right in (e)) does not reach the 
$\theta$-axis and the number of arches is still four. 
The reconstruction process was studied 
well in Paper I. However, we have now found that this is promoted by the 
periodic orbits. 
Again in (e) $a=-0.186$, the unstable periodic points marked as five 
open circles stay at the vertices of the Schubart region.
On the other hand, the stable periodic points marked as remaining 
five open circles are on the midway to the foot-points, although the
stable region is no longer visible.
The non-dominant orbits with $\nu_\infty=7/11$ have moderate distance 
from the Schubart orbit at (f) $a=-0.210$ and arrive at the boundary 
of the Schubart region at (g) $a=-0.230$. Since the stable regions
are small, they are less effective to the shape of the Schubart region 
than the dominant periodic orbits. 
The number of arches is five already in (f) $a=-0.210$. According to 
Paper I, the number of arches becomes five at the totally degenerate
case ($a \fallingdotseq -0.208)$ when the germs $\reg{5,\cdots}$ touch the 
$\theta$-axis. It is important to note that the value of $a$ such that 
the dominant periodic orbits bifurcate is far from that of total degeneracy.
The next dominant orbits with $\nu_\infty=4/6$ exist at (h) $a=-0.250$. 
In this case, since there is two stable orbits, while their period 
is three, the number of islands is six. This makes, as a result, the 
separatrix be of hexagram shape. The periodic points with $\nu_\infty=3/5$ 
still exist at the place marked with $u_i$.

\begin{figure}[htbp]
  \begin{center}
  \pit{a}{-0.150,0}{a-015.\EXT}
  \pit{b}{-0.160,0}{rg_a-0160b0.\EXT}
  \pit{c}{-0.166,0}{rg_a-0166b0.\EXT}
  \end{center}
  \caption{The periodic points and the separatrix on the Poincar\'e section for $\nu=3/5$}
  \label{fig280}
\end{figure}
\backfigure
\begin{figure}[htbp]
  \begin{center}
  \pit{d}{-0.170,0}{rg_a-0170b0.\EXT}
  \pit{e}{-0.186,0}{rg_a-0186b0.\EXT}
  \pit{f}{-0.210,0}{a-021b0.\EXT}
  \end{center}
  \caption{(continue)}
  \label{fig285}
\end{figure}
\backfigure
\begin{figure}[htbp]
  \begin{center}
  \pit{g}{-0.230,0}{a-023b0.\EXT}
  \pit{h}{-0.2500,0}{a-025b0.\EXT}
  \end{center}
  \caption{(continue)}
  \label{fig285}
\end{figure}

\section{Summary}

We have studied the movement of the periodic points bifurcated from 
the fixed point (the Schubart orbit) and their influence on the 
structure of the Poincar\'e section for symmetric mass configuration. 
The following is the summary of the present paper.  
\begin{itemize}
\item The periodic orbits with $\nu=(n-2)/n$, $n$ being integer, 
are influential on the structure of the Poincar\'e section. 
There is a rule about the number of orbits for these type orbits. 
A pair of stable and unstable orbits with period $n$ bifurcate for 
odd $n$, while two pairs bifurcate for even $n$. 

\item There is a value $a=a_{1/3}$ where the rotation number $\nu$ 
at the fixed point is 1/3. Whether $a$ is less or greater than $a_{1/3}$, 
as $|a - a_{1/3}|$ increases, the periodic orbits bifurcate from the 
Schubart orbit one after another. As a result, there are no periodic 
orbits bifurcated from the Schubart orbit at $a=a_{1/3}$. 

\item While the periodic points are inside the Schubart region, the germs,
which bifurcate from arches grow along the separatrix of polygramic 
shape.  The Schubart region takes also polygram shape.
After the periodic points go out of the Schubart region, the Schubart 
region becomes polygonal. The germs grow along this polygon.

\item The stable periodic points leave the fixed point quickly. 
They collect germs and sink toward the $\theta$-axis. 
This collection of germs results in re-composition of arches.
On the other hand, the unstable periodic points stay at the 
vertices of the Schubart region for long time.

\item At the moment of the bifurcation of the dominant periodic orbit 
for $\nu_\infty=(n-2)/n$, the number of arches is still $n-1$. 
\end{itemize}
Several problems are remained unsolved. First, we could not determine
where the stable periodic points tend to. One possible answer is to reach 
the foot-points. Second, the place where the unstable periodic points should 
finally go is also unclear. Third, we have found that 
two stable orbits bifurcate, when $(n-2)/n$ is divisible by 2. 
The possibility that three or more (non-dominant) periodic orbits 
bifurcate is unknown.  Fourth, we have not considered the 
periodic orbits which do not bifurcate from the Schubart orbit. 
For example, in Fig. \ref{fig280} (a) $a=-0.15$, there are four 
black regions inside arches (one region for one arch). We have 
confirmed that there are two period-2 orbits inside these regions 
such that one has its orbital point in the first and the third 
arches and the other in the second and the fourth arches.
These periodic orbits has a different symbol sequence, $(21121)^\infty$, 
from that of the Schubart orbit.

\def\Bibitem#1#2{\bibitem[#1]{#2}}

\end{document}